RESEARCH ARTICLE

# manylp: reproducible, exact, and batched linear programming for genome-scale dynamic flux balance analysis

Fabio Cumbo [1,*]
[1] Computational Life Sciences, Cleveland Clinic Research, Cleveland Clinic, Cleveland, OH 44195, USA

* To whom correspondence should be addressed:
Fabio Cumbo[1], Computational Life Sciences, Cleveland Clinic Research, Cleveland Clinic, 9500 Euclid Avenue, NA2, Cleveland, OH 44195, USA
Email: cumbof@ccf.org

## ABSTRACT

Dynamic flux balance analysis (dFBA), alongside its spatial, community, and ensemble variants, is a foundational tool for predicting microbial metabolism and ecosystem dynamics. These predictive models generate a high computational burden, often requiring the solution of $10^6$ to $10^8$ small linear programs (LPs) per study. Because these LPs represent identical metabolic networks under shifting environmental conditions, they share a common constraint matrix and objective, differing only in their variable bounds. General-purpose CPU solvers (e.g., dual simplex or interior point) do not exploit this temporal and ensemble coherence, treating each LP independently and paying factorization costs on every solve. Conversely, modern matrix-free GPU solvers, which utilize primal-dual hybrid gradient (PDHG) methods, are designed for single large-scale problems at moderate accuracy and often fail to converge on the degenerate, badly scaled matrices typical of genome-scale FBA. Beyond performance bottlenecks, these solvers introduce an issue of reproducibility: because metabolic optima typically form high-dimensional faces rather than unique vertices, solvers silently return arbitrary flux vectors. Consequently, the driving exchange fluxes of the ODEs, and thus the resulting biological predictions, are ambiguous and depend on the solver's implementation and pivoting rules.

To resolve this dual challenge, we propose manylp, a bounds-parametric batched LP solver that returns certified, vertex-exact solutions through a certify-and-repair architecture. Building upon the classical bunching techniques of stochastic programming, manylp leverages the fact that an optimal basis remains dual feasible when only bounds change, allowing the solution to be expressed as an affine function of those bounds. The solver caches optimal bases and certifies entire batches of LPs simultaneously via a single dense matrix product per basis, utilizing lazily computed affine laws, fused multi-threaded CPU kernels, or GPU acceleration. Uncertified LPs are sampled using farthest-point selection and re-solved via a warm-started lexicographic dual simplex, with new bases propagated across the batch until a yield guard triggers direct mode. manylp verifies primal feasibility, lexicographic optimality, uniqueness, and Farkas infeasibility per LP using independent double-precision arithmetic. Furthermore, manylp introduces persistent basis atlases, enabling the caching of metabolic regimes across different runs to eliminate redundant simplex solves for previously explored biological states.

Benchmarking against 25 configurations of 10 exact and first-order solver packages highlights substantial throughput improvements. On a recorded, temporally coherent genome-scale workload, the CPU implementation of manylp certified 47,889 LPs per second, operating 12 to 33 times faster than HiGHS, Gurobi, FICO Xpress, GLPK, and OR-Tools GLOP. On a single NVIDIA A100 GPU, community dFBA ensembles scaled to 682,000 LPs per second, requiring 76 simplex solves to evaluate 94.6 million LPs. First-order GPU competitors (NVIDIA cuOpt, MPAX) were orders of magnitude slower, completing under 1-8% of the workload within a 15-minute budget. While algorithms like cuOpt PDLP successfully converged on the tiny fraction of LPs they processed, solvers such as MPAX and batched PDHG largely failed to converge to the required accuracy. When integrated into the μODE simulator, manylp accelerated single biological trajectories 89-fold, 16-member ensembles 635-fold, and spatial grids 100-fold compared to legacy backends. This microsecond-level LP evaluation makes higher-order embedded Runge-Kutta (RK45) adaptive integration computationally affordable, reducing biomass integration errors 83-fold over standard fixed-step Euler schemes. The method also outperformed trajectory-based event simulators like dfba, which proved 4.6 to 6.6 times slower per trajectory on diauxic shifts and failed on genome-scale models.

Finally, manylp measures model uncertainty by enforcing a lexicographic rule, which ensures reproducible fluxes across solvers. We computed an envelope of 49 discrete flux-selection policies in a single batched simulation, revealing that mathematically equivalent optima alter predicted acetate concentrations 4.2-fold, butyrate 3-fold, and shift the estimated relative abundance of simulated species. Beyond microbiology, manylp generalizes to Netlib LPs subjected to right-hand-side uncertainty, outperforming general-purpose solvers by up to 291-fold. By decoupling numerical certification from exploratory optimization, manylp redefines the computational limits and reliability of dynamic flux balance analysis.

## MAIN

Flux balance analysis (FBA) is the mathematical and computational cornerstone for predicting the metabolic states of cellular systems [1–3]. The method operates by resolving a linear program (LP) over a genome-scale metabolic network (GEM): it maximizes a defined biological objective, typically biomass production, subject to the constraints of steady-state mass balance and specific flux bounds. Dynamic flux balance analysis (dFBA) elevates this paradigm by coupling these static metabolic LPs to ordinary differential equations (ODEs) that explain the concentrations of extracellular metabolites and biomass over time [4,5]. Under the standard static optimization approach, a simulator must advance time by discretely resolving one LP per species at every integration step, with the metabolic uptake bounds dynamically updated to reflect current environmental concentrations [5–7].

As modern computational biology scales from single-species bioreactors to spatially resolved microbial communities, this requirement generates a massive computational bottleneck [8–10]. Spatial and community simulators are particularly demanding. For instance, COMETS resolves one LP per microbial species per spatial grid cell at every step [11–13], while the μODE [14] engine integrates complex, multi-species communities reconstructed directly from metagenome-assembled genomes [15,16]. The burden is multiplied further by the necessity of parameter sweeps, Monte Carlo ensembles modeling dietary and kinetic variations, and rigorous uncertainty analyses [17]. Consequently, a single comprehensive systems biology study easily requires the evaluation of $10^6$ to $10^8$ independent LPs.

Despite this staggering volume, these LPs exhibit a rigid, highly exploitable mathematical structure: they are relatively small, typically containing $10^3$ to $10^4$ variables, and all LPs generated for a single species share the exact same stoichiometric matrix and objective function [18–20]. They differ only in a narrow subset of exchange variable bounds. Furthermore, these shifting bounds exhibit profound temporal coherence (changing only marginally from one integration time step to the next) and ensemble coherence (varying slightly across parallel simulation members).

Current general-purpose LP solvers utilize this inherent biological structure poorly. Classical CPU-based algorithms, whether utilizing the dual simplex or interior point methods, are architected to solve one LP at a time [21–24]. While they attempt to exploit temporal coherence through warm starts, they still incur severe computational overhead by executing matrix factorization and pricing operations during every independent solve.

Recent advancements in GPU-accelerated optimization have not resolved this specific bottleneck. Modern GPU LP solvers, largely based on the primal-dual hybrid gradient (PDHG) method (e.g., PDLP, cuPDLP, cuOpt, and MPAX), target single, large-scale problems at moderate accuracy [25–28]. First-order methods like these struggle with the degenerate, ill-conditioned matrices typical of genome-scale FBA, resulting in slow convergence or complete failures [29,30]. Similarly, batched optimization layers such as OptNet, cvxpylayers, and OSQP [31–35] are designed to tackle small problems up to ADMM or interior-point accuracy, lacking the capacity to reliably yield the exact vertex solutions required by FBA.

Beyond pure computational throughput, these bounds-parametric LPs mask a deeper scientific vulnerability. While FBA successfully pins the optimal maximal growth rate of a cell, it generally fails to uniquely identify the internal flux vector. The biological optimum is typically a high-dimensional face of the feasible polytope rather than a single, unique vertex [36,37]. Because the exchange fluxes that drive the dFBA differential equations are extracted directly from whichever optimal vertex the solver happens to return, the resulting biological trajectories depend on the choice of solver and its internal pivoting rules [38,39].

The biological impact of this solver ambiguity can be vast, particularly in models of the human gut microbiome. Our driving application in this study is a community model of the human gut: twelve genome-scale metabolic reconstructions of prevalent gut species [16] (including *Bacteroides thetaiotaomicron*, *Bacteroides fragilis*, *Faecalibacterium prausnitzii*, *Roseburia intestinalis*, *Eubacterium rectale*, *Segatella copri* [40], and *Akkermansia muciniphila*) growing together on a Western diet [41] for 48 hours. The quantities such simulations are used to predict are the short-chain fatty acids (acetate, propionate, butyrate) and gases the community releases, and how its composition shifts, because these mediate most of the microbiome's effects on the host [42]. Acetate is the central currency of this network: it is secreted by saccharolytic *Bacteroides* and *Bifidobacterium* species and consumed by butyrate producers [43–45] such as *F. prausnitzii* and *R. intestinalis*. A prediction of acetate is therefore also a prediction of cross-feeding, and of butyrate.

Because a species can achieve the same optimal growth rate while secreting drastically different amounts of acetate, the arbitrary vertex chosen by the solver directly dictates the availability of these critical nutrients. This divergence alters what neighboring species can consume, allowing initial solver-induced variations to persist and compound over time, fundamentally shifting the predicted community composition and total metabolic output [46,47]. To complement this complex network dynamics test, we also use the classical *E. coli* diauxic shift reference [5,48,49]. Its sharp metabolic switch serves as a standard stress test for evaluating dFBA adaptive integrators.

Historically, the community has utilized flux variability analysis (FVA) to measure the extent of these alternative optima, or mixed-integer methods to enumerate the vertices [36,37]. To enforce a unique solution for a single trajectory, various rules have been proposed, including parsimonious FBA (pFBA), geometric FBA, lexicographic priority lists, L2 regularization (as seen in MICOM), or the application of secondary objectives [7,47,50–54]. However, in dFBA, handling non-uniqueness remains fiercely debated. While some studies select a single trajectory calibrated to empirical data [55], others propagate several trajectories generated via varying regularization weights, treating them as scenarios for robust model-predictive control [56,57]. The biological impact of this solver ambiguity can be vast: while reordering lexicographic objectives had negligible effects in some single-species spatial models [58], community models of the human gut demonstrate that optimal acetate secretion can span an alarmingly wide range [46,47], drastically altering cross-feeding predictions.

The mathematical foundation for overcoming both the throughput bottleneck and the alternative optima crisis lies in classical parametric programming. As a function of its right-hand side, the solution to an LP is piecewise affine over polyhedral critical regions [59–62], with one region corresponding to each optimal basis. Multiparametric programming attempts to enumerate these regions offline, while explicit model-predictive control evaluates the stored affine law online [61,63].

Stochastic programming has exploited this property without relying on explicit enumeration [64]. Because second-stage LPs differ only in their right-hand sides, an optimal basis for one scenario remains dual feasible for all others [65]. The classical technique of bunching checks primal feasibility for every remaining scenario and utilizes a warm-started dual simplex to solve the rest. However, applying these concepts to genome-scale dFBA requires handling general variable bounds rather than just non-negative right-hand sides, certifying lexicographic degeneracy, caching infeasibility, and scaling the matrix operations for modern hardware architectures. Furthermore, existing batched tools do not reuse bases across instances or provide rigorous mathematical certificates for their answers.

Several methods already reuse optimal bases or critical regions, but each covers only part of what dFBA ensembles require. Ji et al. (2017) generate critical regions on demand during Monte Carlo simulation of power-system dispatch, solving a new optimization problem only when a sampled scenario falls outside every region found so far [66]. Akbari and Barton (2018) apply multiparametric programming to FBA itself and compute the critical regions of a metabolic network explicitly, an approach whose cost in general grows rapidly with the number of varying bounds [62]. Bertsimas and Stellato (2021), as well as Misra et al. (2022), show that learning approaches predict the optimal active set of a new instance instead, but the prediction is not a certificate and must still be checked or repaired by a solver [67,68]. Within dFBA, Höffner et al. (2013) and Brunner and Chia (2020) keep the optimal basis along a trajectory and re-solve the LP only when that basis stops being valid, which greatly reduces the number of LPs per trajectory but still integrates each trajectory separately [10,53]. On GPUs, Gurung and Ray (2019) run a separate simplex method for each LP of a batch [35], and Blin et al. (2026) solve batches of LPs with first-order methods that return ε-approximate rather than vertex solutions [33]; neither shares optimal bases between the LPs of a batch. No existing method combines on-demand critical regions shared across an entire batch of species, time steps and ensemble members, batched evaluation on CPUs and GPUs, and per-LP certificates of optimality, uniqueness, and infeasibility.

To bridge this gap, we propose manylp, a batched LP solver engineered specifically for the bounds-parametric LP families generated by dFBA. manylp translates the classical bunching concept into a highly scalable, certified solver tailored for modern dFBA workflow. Our primary contributions are:

1. manylp maintains a dynamic pool of cached optimal bases that certify entire batches of LPs using a single dense matrix product per basis, executing on GPUs or through fused multi-threaded CPU kernels. The certificate covers lexicographic optimality, uniqueness of the optimum, and infeasibility through Farkas rays [69]. LPs falling outside cached regions are resolved by a warm-started dual simplex, producing vertex solutions whose validity is rigorously checked in double-precision arithmetic independently of the simplex solver.

2. Affine laws are computed lazily, restricting operations exclusively to parametric variables that actively shift. Progressive repair algorithms propagate newly discovered bases to the entire batch, while a yield guard bounds the worst-case performance on highly incoherent batches. Crucially, persistent basis atlases carry cached bases across independent simulations; under a unique flux rule, these atlases can never alter a biological result.

3. manylp extends scenario trajectories by evaluating dozens of systematic, reproducible selection policies over the optimal face in lockstep. This generates an envelope of whole simulations, effectively quantifying how much of a dFBA prediction the model itself leaves undetermined, a vital metric that conventional solvers silently collapse into a single arbitrary point.

4. We integrated manylp into the μODE community simulator and conducted an exhaustive audit of every returned solution of 11 solvers across a panel of 25 different configurations (including HiGHS, Gurobi, FICO Xpress, GLPK, OR-Tools, OSQP, NVIDIA cuOpt, and MPAX) [23,28,34]. The evaluation encompasses genome-scale workloads, Netlib LPs under right-hand-side uncertainty [70], models spanning 95 to 10,600 reactions, and direct comparisons against classical bunching, native lexicographic modes, exact rational solving [71,72], and alternative dFBA trajectory simulators [10,73].

## RESULTS

We evaluated manylp against 25 configurations of 10 state-of-the-art exact and first-order solver packages across a range of workloads (Supplementary Tables S1 and S2), spanning recorded genome-scale temporal dynamics, end-to-end community spatial simulations, and generic linear systems under right-hand-side uncertainty. The findings demonstrate that decoupling rigorous mathematical certification from exploratory simplex optimization yields massive throughput gains. Furthermore, we show that this microsecond-level computational efficiency enables high-order adaptive integration and systematically resolves the biological ambiguity inherent to alternative optimal states in dFBA.

**A human gut community as the driving use case**

Every biological workload evaluated in this study derives from either a 12-species human gut community model or a classic *E. coli* core diauxic shift. The gut community integrates genome-scale metabolic reconstructions growing on a simulated Western diet. To benchmark solver performance across diverse operational extremes, we derived specific simulation workloads from these baselines:

- <u>Coherent workload:</u> 185,088 LPs recorded sequentially along the first 24 hours of the gut community trajectory, representing the highly coherent access pattern of standard dFBA.
- <u>Snapshot workload:</u> 23,808 LPs generated by sampling the community trajectory every 1.6 hours (every 16th time step), serving as a stress test that removes temporal coherence to force solver cold starts.
- <u>Ensembles:</u> parallel simulation members generated by applying log-normal perturbations to the dietary input bounds, testing structural reuse across slightly varying environments.
- <u>*E. coli* diauxie:</u> a 12-hour single-species simulation of glucose consumption and subsequent acetate reuse, providing a standard reference for adaptive integration across sharp metabolic switches.

**Batched certification delivers order-of-magnitude throughput gains**

To establish baseline performance on real-world biological access patterns, we first benchmarked solvers on a recorded, temporally coherent dFBA workload comprising 185,088 genome-scale LPs. On this workload, manylp achieved a steady-state throughput of 47,889 LPs per second on a single 16-thread process. This performance establishes a 12- to 33-fold speedup over state-of-the-art general-purpose solvers running across 32 worker processes, including warm-started HiGHS (12.0x), Gurobi (16.2x), FICO Xpress (17.7x), GLPK (26.1x), and OR-Tools GLOP (32.7x) [23].

The true capacity of the certify-and-repair architecture is realized as ensemble size scales. In community dFBA experiments tracking a 12-species microbiome over 48 hours, we scaled the ensemble (E) from a single trajectory to 16,384 diet-perturbed members. Batching efficiency allowed the GPU implementation to overtake the CPU starting at E=256. At the maximum scale, evaluating 94.6 million LPs required merely 76 simplex solves, as the vast majority of biological states fell into shared optimal critical regions. Consequently, the GPU scaled to 682,000 LPs per second, an 8.6-fold increase over the CPU's plateau (Figure 1). These throughput levels are achieved while returning certified, vertex-exact answers. In contrast, alternative GPU solvers (NVIDIA cuOpt, MPAX) were orders of magnitude slower than manylp on the identical NVIDIA A100 GPU, completing under 1-8% of the coherent workload within a 15-minute budget. Their limitations varied strictly by algorithm: while cuOpt PDLP successfully converged to $10^{-6}$ accuracy on the fraction of the workload it reached (0.24%), solvers like MPAX, cuOpt barrier with crossover, and our custom batched PDHG [26–28] overwhelmingly failed to converge, solving between 0% and 12% of the LPs they processed (Table 1).

To define the boundary conditions of the solver, we subjected manylp to a weakly coherent snapshot stress-test workload consisting of 23,808 LPs (Supplementary Table S2). On this workload, classical bunching utilizing 8 processes achieved 4,414 LP/s [65], outpacing the plain-FBA manylp CPU implementation (3,477 LP/s) and operating at roughly 0.9x the speed of the GPU implementation (4,930 LP/s). However, manylp's most significant advantage emerges under the strict pFBA-unique rule: when generating certified unique fluxes, manylp sustained 4,860 LP/s, outperforming the equivalent HiGHS lexicographic implementation (732 LP/s) [23,74] by 6.6-fold.

**Resolving the alternative optima crisis in microbiome modeling**

While exact solvers mathematically agree on the optimal biological growth rate for a given time step, they routinely disagree on the specific flux distribution required to achieve it. Because the biological optimum typically forms a high-dimensional face rather than a single vertex, exact solvers return divergent, equally valid exchange fluxes [36] that differ by up to 2.7 mmol $gDW^{-1}$ $h^{-1}$ for the exact same LP.

To quantify the downstream biological consequences of this solver-induced ambiguity, we integrated a 12-species gut community model over 48 hours using an alternative-optima envelope. This approach evaluated 49 distinct flux-selection policies simultaneously (including the pFBA-unique rule, 32 random directions on the optimal face, and extreme secretion profiles for eight fermentation products) in a single batched run. All 282,828 LPs generated during this envelope simulation were computationally certified as optimal and unique under their respective policies.

The results reveal that mathematically equivalent FBA optima dictate profoundly divergent biological predictions (Figure 2, Table 2). Across the 49 policies, predicted short-chain fatty acid concentrations varied massively: acetate exhibited a 4.2-fold divergence (ranging from 0.209 to 0.883 mM), while butyrate varied 3.0-fold (0.115 to 0.343 mM). Furthermore, this uncertainty propagated beyond metabolite pools into the fundamental community composition. As the cross-feeding metabolite availability shifted across different solver trajectories, the final relative abundance of keystone species such as *B. fragilis* fluctuated between 15.7% and 20.6%. Even within a single, conventional exact solver configuration (e.g., COBRApy with GLPK [75]), the arbitrary vertex selected by the solver resulted in final biomass differences of up to 1.4% when compared to a systematic uniqueness rule.

manylp resolves this reproducibility crisis by strictly enforcing a pFBA-unique lexicographic rule, which isolates a single, geometrically consistent vertex. This ensures deterministic, platform-independent trajectories. When benchmarking 48-hour biological simulations, manylp's trajectories matched an independent HiGHS implementation applying the same rule to within $4\times10^{-15}$ gDW $L^{-1}$ for biomass and $9\times10^{-13}$ mM for extracellular metabolites.

Finally, manylp ensures these trajectories are numerically safe. To verify the internal double-precision certificates, we sampled 200 LPs and solved them using SoPlex in exact rational mode [71,72], confirming that manylp's certified optima perfectly matched rational arithmetic to within $1.7\times10^{-11}$. In contrast, commercial solvers proved alarmingly unreliable when tasked with this level of lexicographic precision natively: running FICO Xpress under its default settings resulted in non-lexicographically optimal fluxes for 29% of the tested genome-scale LPs, a silent mathematical failure entirely caught by manylp's per-LP certification.

**End-to-end acceleration of community and spatial simulations**

To measure the practical impact of batched certification within a production environment, we integrated manylp as the core optimization backend for the µODE simulator. This allowed us to benchmark end-to-end biological simulations, inclusive of all kinetic evaluations, ODE updates, and internal bookkeeping, against µODE's per-LP backend [14] (COBRApy coupled with GLPK [75]).

The integration of manylp yielded massive wall-clock reductions for both single-trajectory and lockstep ensemble simulations tracking a 12-species gut community over 48 hours (Table 3). For a single biological trajectory, evaluating 5,772 LPs dropped from 11 minutes (653.6 seconds) to just 7.7 seconds, an 89-fold acceleration. The gains compound aggressively as ensemble size (E) increases: simulating a 16-member diet-perturbed ensemble accelerated 635-fold, dropping from over 3 hours (10,920 seconds) to 17.2 seconds on the GPU. Strikingly, a 1,024-member uncertainty ensemble, requiring the resolution of nearly 5.9 million LPs, completed in just 13 minutes (776.1 seconds). Executing this exact experiment using the original solver path would take approximately one week.

Because ensemble members share the vast majority of their critical regions, manylp required only 72 simplex solves to discover the 30 distinct critical regions that certify the entire 1,024-member ensemble. Beyond these 30 regions, over 5.9 million LPs were evaluated purely through matrix certification from the basis pool. At this scale, the LP optimization is no longer the computational bottleneck: while the end-to-end 1,024-member simulation took 776 seconds, manylp's standalone certification required only 23 seconds for the identical LPs, indicating that over 95% of the total runtime is now consumed by the simulator's native Python engine overhead.

We observed comparable throughput leaps in spatial FBA configurations. Using μODE's COMETS-style spatial engine [11,12], which solves one LP per species per grid cell per step, manylp batches all cells of a step together for each respective organism. On an 8x8 spatial grid simulating 1 hour of diffusion dynamics, manylp accelerated the integration 100-fold compared to the historical backend (8.7 seconds versus 868.5 seconds). As grid resolution scales, neighboring cells share critical regions much like ensemble members do, driving extrapolated speedups to 206-fold on a 16x16 grid and 274-fold on a 32x32 grid.

**Enabling higher-order adaptive integration**

Historically, dFBA simulators have relied on simple, fixed-step explicit Euler integration schemes [5,10,53] because the computational cost of evaluating the LP at intermediate stage points was prohibitive. Because manylp reduces the cost of a certified LP evaluation to the microsecond regime, higher-order adaptive integration methods, such as embedded Runge-Kutta (RK45) [76,77], become computationally affordable for routine genome-scale modeling.

We compared μODE's standard production scheme (explicit Euler at a fixed time step of 0.1 hours) against an RK45 adaptive integrator on both the 12-species gut community (over 48 hours) and an *E. coli* core diauxic shift model (glucose to acetate, over 12 hours). Errors were measured as the maximum relative deviation against an exceptionally fine reference Euler simulation (dt=$10^{-4}$ hours for diauxie and $10^{-3}$ hours for the gut community), which was only possible to compute because nearly every LP was a certified cache hit.

The integration of RK45 yielded massive accuracy improvements (Table 4). On the 48-hour gut community model, utilizing RK45 at a relative tolerance (rtol) of $10^{-4}$ reduced the maximum relative biomass error 83-fold compared to the production Euler scheme. This precision required only 2.75 times the number of LP evaluations (15,864 LPs versus 5,772 LPs).

The adaptive integrator is especially critical during rapid metabolic transitions. During the *E. coli* diauxic shift, the standard fixed-step Euler scheme (dt=0.1h) yielded a relative error of 6.5 in predicted glucose and acetate concentrations. Even when the Euler step size was reduced 10-fold (dt=0.01h), the RK45 integrator (rtol $10^{-6}$) proved 97 times more accurate in tracking biomass while actually requiring fewer total LP evaluations (1,408 versus 2,402).

Finally, we explicitly evaluated the utility of regime-event splitting, an algorithmic technique designed to precisely locate the exact timestamps where uptake bounds shift and bind [53,78]. While manylp successfully located eight regime switches during the diauxic shift at rtol $10^{-6}$, the technique did not meaningfully improve overall integration accuracy. Instead, the event-location overhead increased the required LP evaluations by 84% (from 1,408 to 2,586) and more than doubled the integration wall time (from 0.77 to 1.96 seconds). The accuracy gains observed in our benchmarks stem entirely from the adaptive step sizing, demonstrating that explicit regime-event tracking is an unnecessary computational burden given manylp's evaluation speeds.

### Algorithmic optimization: progressive repair, basis atlases, and sparsity

While manylp achieves unprecedented steady-state throughput via cache hits, initiating a novel simulation presents a cold-start penalty where the basis pool is entirely empty. To mitigate this, we engineered three optimization layers (progressive repair, persistent basis atlases, and lazy affine law evaluation) that collectively minimize the computational footprint of simplex solvers and device memory.

We first conducted a cold-start ablation study on the 12-species community model to isolate the cost of warming up the cache (Table 5). Under a naive eager repair strategy, the engine attempts to solve every uncertified ensemble member simultaneously. By implementing progressive repair, where representative LPs are iteratively chosen via farthest-point sampling [79] in normalized bound space and newly discovered bases are immediately propagated across the batch, we observed an 11-fold reduction in required simplex solves (dropping from 864 to 76 for both 64- and 1,024-member ensembles). This intelligent sampling reduced active repair time from 13-17 seconds down to 3.6-6.1 seconds, cutting total solve time by nearly half (1.8-1.9x) for a 64-member batch. It is important to note that progressive repair and atlases introduce Python dispatch and management overhead; on the CPU at E=1,024, eager repair actually completed the batch slightly faster (70 seconds) than progressive repair (79 seconds) and the atlas (75 seconds). However, on the GPU, where matrix multiplication operations dominate and throughput scales massively, progressive repair and atlases maintain a strict advantage.

To bypass the cold-start penalty entirely across independent studies, manylp introduces a persistent basis atlas. Because FBA critical regions depend strictly on the network topology and objective, a pool of certified bases can be saved and reused. When initiating a simulation of a completely unseen diet ensemble using a pre-computed basis atlas, simplex solves were eliminated entirely. Repair time fell to a negligible 20-40 milliseconds. In practice, this enables a paradigm shift in open science: laboratories can distribute lightweight basis atlases alongside their community models, ensuring that downstream uses never incur simplex optimization costs for biological regimes that have been previously explored.

Finally, we optimized the memory footprint of the cached bases by exploiting biological limiting-nutrient sparsity. A given basis's affine law mathematically requires only the parametric columns that are nonbasic at a bound. Because microbial growth is typically limited by only a handful of nutrients simultaneously, the vast majority of exchange bounds do not actively constrain the solution at any one time. By evaluating a lazy parameter-sparse affine law, we found that only 21.4 out of 234 nonbasic parametric columns (9.1%) were ever activated per basis on average. This sparsity reduced the required GPU device memory for 33 certified bases 7.5-fold (from 87.1 MiB down to 11.6 MiB) without negatively impacting throughput. At a fixed memory budget, this allows the manylp pool to cache 7.5 times more critical regions, a crucial scaling factor for massive genome-scale models featuring thousands of exchange reactions.

### Generalization and boundary conditions

To test the algorithmic boundaries of manylp, we benchmarked it against conceptually related baseline methods (i.e., classical bunching and trajectory-based event simulators) and evaluated its scalability across varying model sizes and non-biological domains.

**Comparisons to classical and trajectory-based baselines –** The certify-and-repair architecture builds upon the classical bunching method of stochastic programming, which tests primal feasibility of cached bases using sparse LU factors [65]. While our optimized bunching implementation captured reuse efficiently on coherent workloads (requiring 279 simplex solves across 8 processes for 185,088 LPs), manylp proved 6.8-fold faster on the CPU. This speedup is driven by manylp's parameterized, lazy affine laws and dense matrix products, which bypass the expensive sparse triangular solves required by classical bunching. Furthermore, traditional bunching cannot cache Farkas infeasibility certificates or certify lexicographic uniqueness.

We also compared manylp against trajectory-based simulators like dfba, which utilizes the Harwood-Barton event method [53,73,74] to minimize the sheer number of LPs solved. On an *E. coli* diauxic shift, dfba required only 17 LPs per trajectory, whereas manylp's batched RK45 integrator evaluated roughly 8,500 LPs per trajectory across a timed 32-member ensemble. However, because manylp reduces the cost per LP to microseconds, it still executed the simulation 4.6 to 6.6 times faster per trajectory despite the drastically higher evaluation count.

Trajectory-based event simulators struggled with stability on larger networks. On the genome-scale iJO1366 model [19] (simulated for 0.5 hours), dfba's event method took 229-495 seconds to complete and yielded an artificially low growth rate (0.62 $h^{-1}$ against the true 0.70 $h^{-1}$), resulting in a 20% biomass error. Furthermore, several dfba runs halted prematurely upon encountering infeasible LPs, whereas manylp correctly navigated these infeasibilities to complete the full integration horizon.

Nevertheless, manylp does possess performance boundaries when compared to trajectory simulators. When evaluating an 8-member iJO1366 ensemble on a single core at an extremely tight adaptive tolerance (rtol $10^{-8}$), manylp required 32.7 seconds per trajectory, ultimately losing to dfba's fallback fixed-interval mode, which completed the integration in 25.6 seconds.

**Algorithmic boundaries: scaling by distinct bases, not model size –** To define when the certify-and-repair paradigm breaks down, we simulated 1,024-member ensembles across eight metabolic models ranging from 95 to 10,600 reactions [16,18–20,49,80–82] (Figure 3A). manylp outperformed 32-process HiGHS on 13 of the 16 tested configurations by 1.9- to 29-fold. The algorithm's advantage was not dictated by the raw size of the model, but exclusively by the number of distinct optimal bases present in the batch. manylp only lost to HiGHS on three highly incoherent batches where the required distinct bases exceeded the solver's default pool capacity of 512 (e.g., the Recon3D model, which features 1,560 independently perturbed exchange reactions).

This fundamental boundary extends universally to non-biological systems (Figure 3B). We evaluated manylp on 44 generic Netlib LPs, perturbing right-hand-side bounds [70,83] to generate 1,024 scenarios per problem (Supplementary Table S3). When the scenarios fell into fewer than 100 distinct bases, manylp proved faster than HiGHS on 20 of 21 problems, achieving up to a 291-fold speedup (geometric mean 12.9x). Against classical bunching on these highly coherent problems, manylp outperformed the best bunching variant on 14 of 21 matrices, yielding a more modest geometric mean speedup of 1.3x.

Conversely, on highly incoherent problems where nearly every scenario required its own distinct basis (100 or more distinct bases), manylp incurred the overhead of repeated simplex solves and certification. In these sparse regimes, both HiGHS and classical bunching proved superior, with bunching operating roughly 1.9x faster than manylp. Therefore, manylp is optimally deployed for massively batched, bounds-parametric workloads (time integration, ensembles, parameter sweeps), whereas conventional simplex codes remain superior for streams of highly incoherent, unrelated LPs.

## DISCUSSION

dFBA requires solvers built for its specific workload: millions of linear programs that share a matrix, differ strictly in their bounds, and fall into a limited number of critical regions. By shifting the computational cost from a per-LP basis to a per-critical-region basis, manylp fundamentally circumvents the bottleneck that cripples general-purpose simplex codes. The certify-and-repair architecture demonstrates that a single warm-started lexicographic simplex solve, when coupled with highly parallelized dense matrix certifications, yields orders-of-magnitude throughput gains for temporally and compositionally coherent biological models.

### The imperative of vertex exactness and reproducibility

The results expose a systemic vulnerability in modern metabolic modeling: solver ambiguity is not a trivial numerical artifact, but a major driver of biological divergence. As demonstrated by the alternative-optima envelope, equally optimal flux choices alter predicted cross-feeding metabolites like acetate by over 4-fold, cascading into profound shifts in predicted community composition. While exact CPU simplex codes reach the optimal face reliably, they return solver-specific vertices [38,39]. Conversely, state-of-the-art GPU first-order methods (PDHG) are architected for massive single LPs and consistently fail to converge to exact vertices on the degenerate [27,29], badly scaled matrices typical of genome-scale FBA. Even native lexicographic modes in commercial solvers occasionally fail silently, returning non-optimal fluxes while reporting successful status flags. By decoupling rigorous double-precision vertex certification from exploratory simplex optimization, manylp provides a mathematically guaranteed pFBA-unique rule. This capability transforms flux uncertainty from an unmeasured artifact into a measurable, controllable variable at the cost of a single batched run.

### Hardware optimization and paradigm shift for open science

manylp efficiently democratizes massive-scale dFBA by operating efficiently across hardware tiers. While GPU acceleration proves transformative for massive ensembles (E>256), reaching up to 682,000 LPs per second, the fused CPU kernels remain highly competitive for smaller batches due to lower kernel launch latencies. Even strictly on a CPU, manylp operates an order of magnitude faster than commercial simplex codes for coherent workloads.

Beyond pure throughput, the introduction of the persistent basis atlas initiates a paradigm shift for open systems biology. Because FBA critical regions depend entirely on network topology and biological objectives, researchers can now compute the metabolic critical regions of a novel community model once and distribute the lightweight atlas alongside the model. Downstream users can entirely bypass simplex optimization costs for regimes already explored, relying strictly on microsecond matrix certification. Under a certified unique flux rule, loading an atlas mathematically guarantees identical, reproducible biological predictions regardless of the user’s local hardware.

### Boundary conditions and limitations

While manylp dominates bounds-parametric, coherent workloads, it is constrained by several specific boundary conditions. As demonstrated by the Netlib stress tests and the highly perturbed Recon3D ensembles, highly incoherent batches, where nearly every LP requires its own distinct optimal basis, neutralize the cache. In these sparse scenarios, the algorithm pays the overhead of Python dispatch, certification, and individual simplex solves, rendering a conventional compiled simplex code the faster alternative. A natural algorithmic extension would be the inclusion of early incoherent detection to hand off such batches entirely to conventional solvers.

Additionally, while manylp's embedded Runge-Kutta integrator drastically improves accuracy, its batched nature currently forces a single, shared step size across all members of an ensemble. At tight error tolerances, the stiffest ensemble member dictates the integration pace for the entire batch, artificially inflating simulation times. Future iterations incorporating per-member step control or stiffness grouping would eliminate this bottleneck [77]. Finally, it must be noted that manylp's certificates are numerical, computed in double precision with tolerances of roughly $10^{-9}$, rather than rigorous interval or rational proofs. While they prove significantly more reliable than standard solver status flags, studies requiring absolute rigorous guarantees must still layer an exact rational check over the returned bases [39,71,84].

## CONCLUSIONS

dFBA does not require a faster general-purpose linear programming solver; it requires a solver fundamentally architected for its specific computational workload, millions of LPs that share a matrix, differ strictly in their bounds, and collapse into a limited number of critical regions. manylp exploits this structural reality through a batched certify-and-repair methodology, yielding certified, geometrically exact vertex solutions an order of magnitude faster than conventional CPU solvers, and several times faster than classical bunching techniques. When scaling to large ensembles on a GPU, these throughput gains extend to multiple orders of magnitude over both exact and first-order competitors.

Crucially, manylp resolves a systemic vulnerability in metabolic systems biology: the silent reproducibility crisis driven by alternative optima. By enforcing a unified, per-LP certified lexicographic rule, the solver guarantees biological trajectories across different environments and platforms. Furthermore, it elevates flux uncertainty from an unmeasured artifact into a measurable, controllable variable. While engineered specifically to eliminate the bottleneck in dynamic microbiome and spatial modeling, the underlying certify-and-repair mechanism is not exclusive to metabolism; it accelerates any scenario analysis or parameter sweep where bounded LPs share critical regions. manylp is fully open source and designed to deploy either as a standalone optimization library or as a drop-in acceleration backend for existing dFBA simulators.

## ACKNOWLEDGMENTS

We acknowledge the use of Google's Gemini 3.1 Pro model to enhance the clarity and readability of this manuscript; this assistance was limited to text editing. We also acknowledge the use of Claude Code powered by Claude Opus 5.5 to assist with the implementation of manylp, its integration into µODE, and the execution of the benchmarks.

The AI played no role in data analysis, research design, or the interpretation of results; all conclusions and insights remain entirely those of the author.

## FUNDING

No funding was received for the conception or writing of this manuscript.

## COMPETING INTERESTS

No conflicts to disclose.

## FIGURES

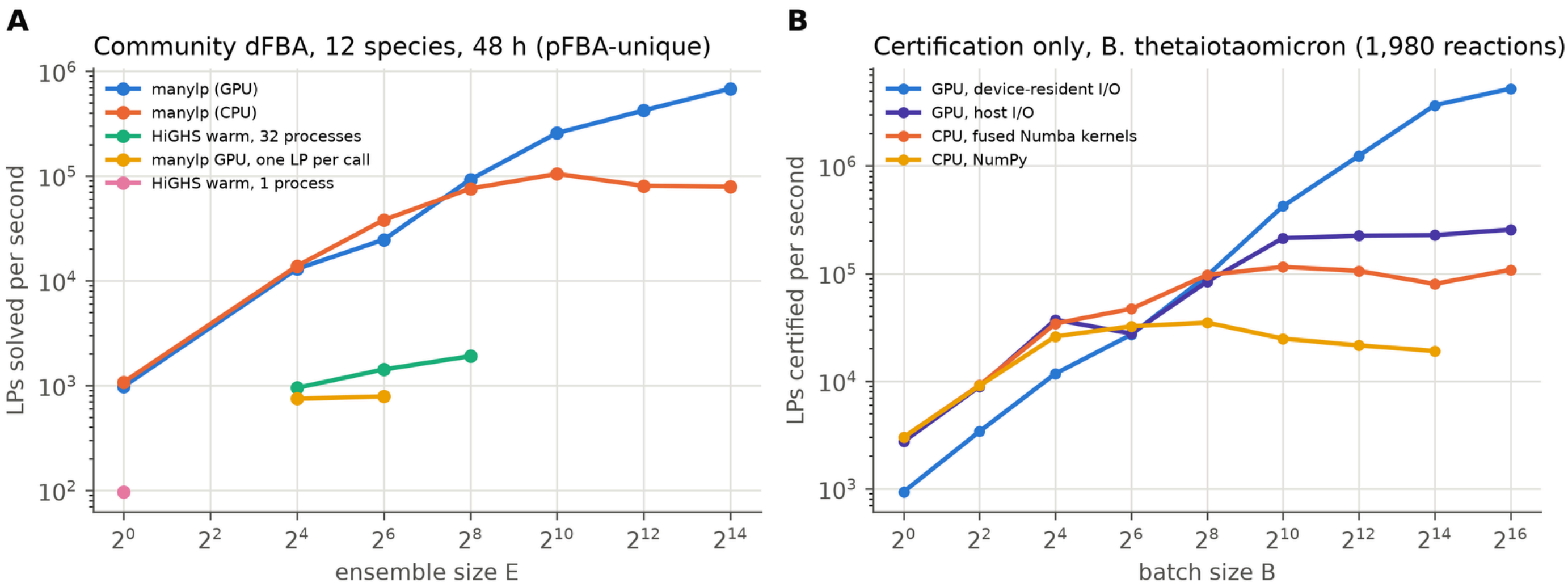

**Figure 1:** Throughput scaling of batched linear programming in community dFBA. Panel A – Steady-state throughput (LPs solved per second) scaling with ensemble size E for a 48-hour 12-species gut community simulation. The GPU implementation of manylp surpasses the CPU plateau at E=256, reaching 682,000 LP/s at maximum scale. Panel B – Certification throughput (excluding model loading and bounding overhead) on a single genome-scale model (*B. thetaiotaomicron*), demonstrating kernel limits where device-resident GPU execution exceeds 5 million certified LPs per second.

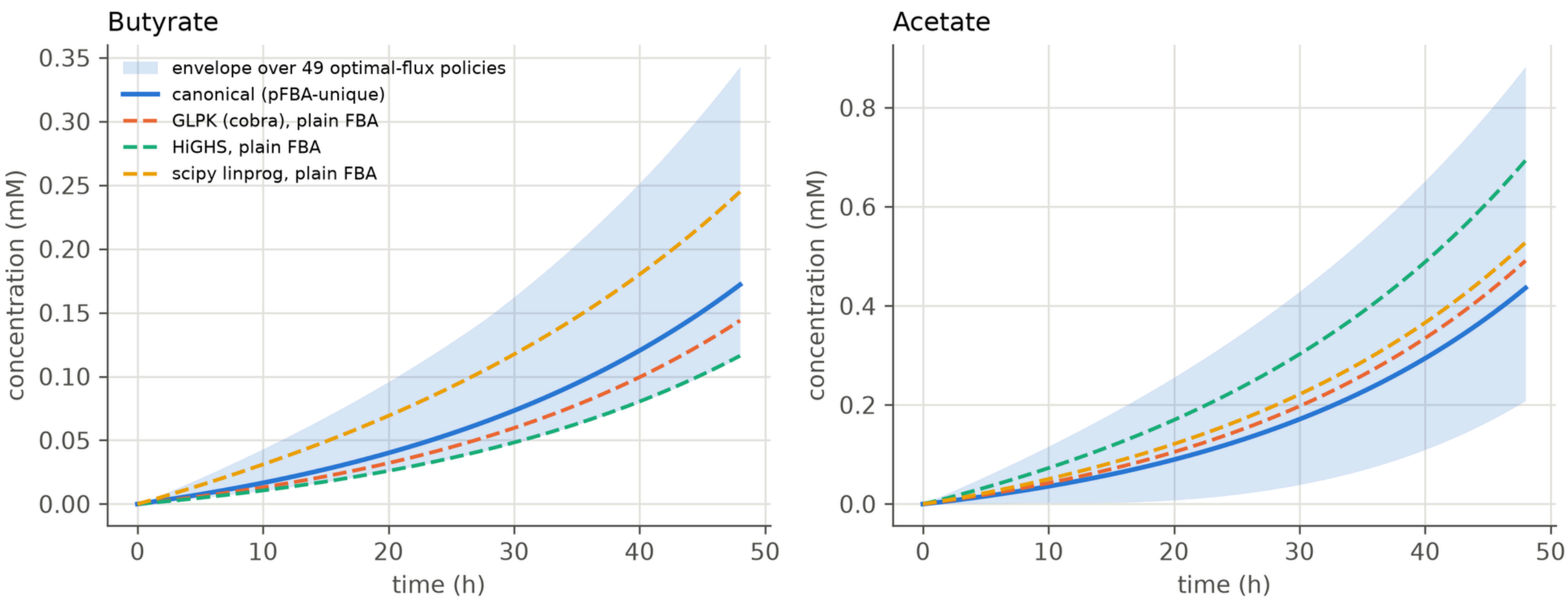

**Figure 2:** Biological divergence driven by alternative optima in dynamic FBA. Simulated concentrations of butyrate (left) and acetate (right) for a 12-species gut community over 48 hours. The shaded band represents the alternative-optima envelope spanning 49 equally valid flux-selection policies, computed simultaneously via a single manylp batched run. Dashed lines illustrate that three conventional exact solvers (GLPK, HiGHS, and SciPy linprog) running standard plain FBA land on entirely divergent, arbitrary trajectories within the envelope. The solid line indicates the reproducible, certified trajectory enforced by the pFBA-unique rule.

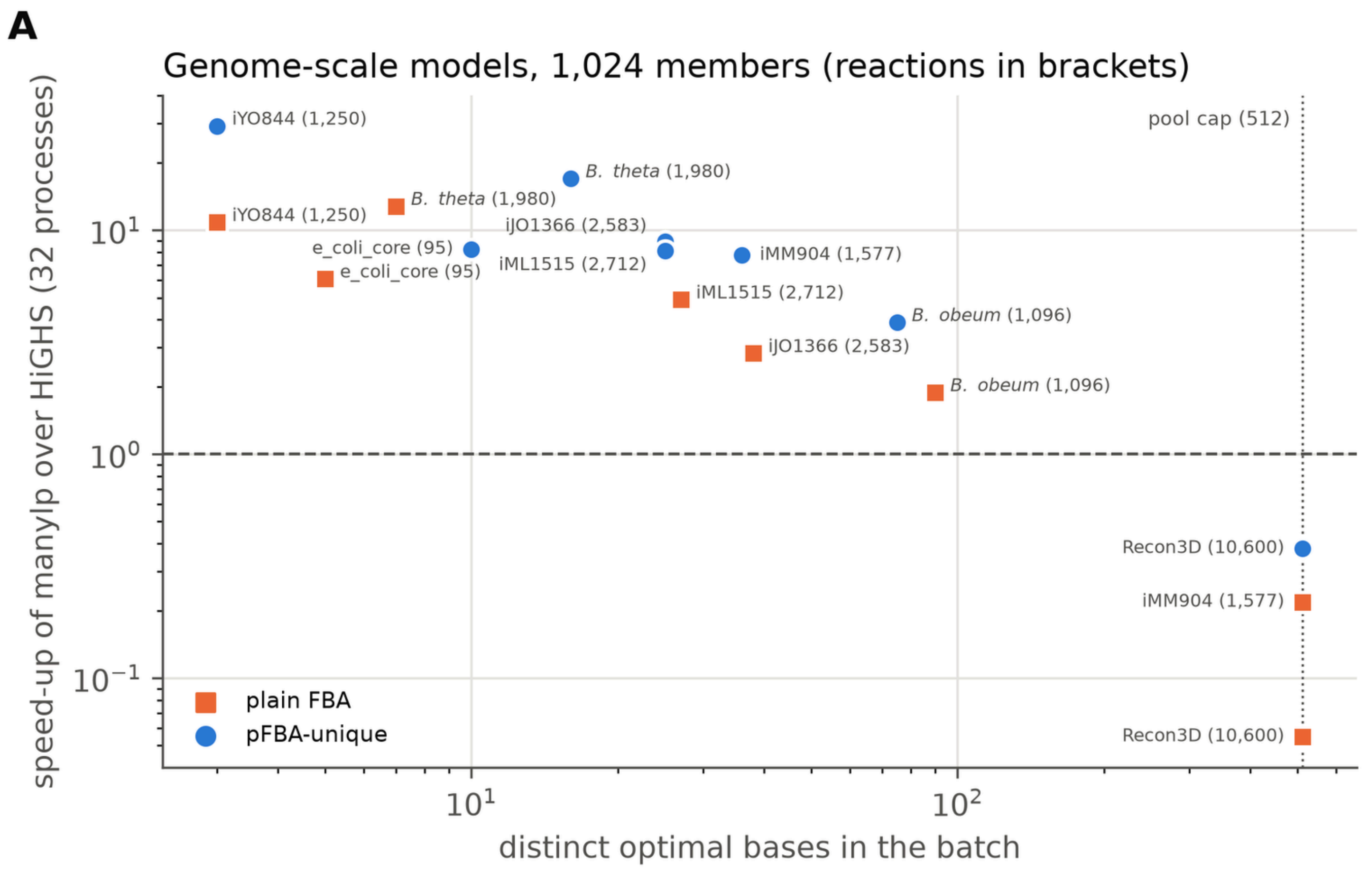

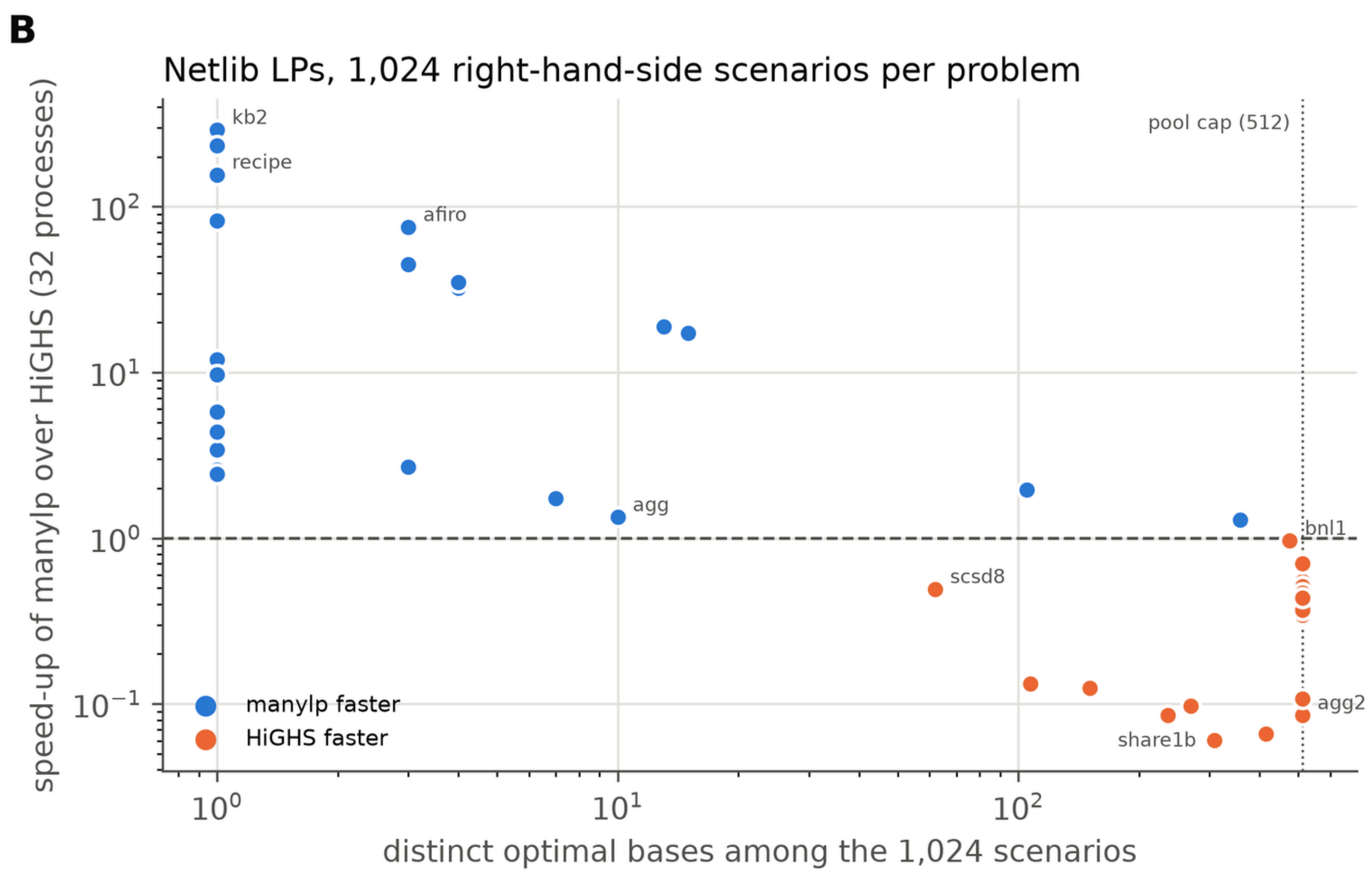

**Figure 3:** Algorithmic boundaries dictated by critical region sparsity rather than model size. Panel A – Speedup of manylp over 32-process HiGHS for genome-scale models ranging from 95 to 10,600 reactions (reaction counts in brackets), plotted against the number of distinct optimal bases required by a 1,024-member batch. Panel B –

Generalization to 44 non-biological Netlib LPs (1,024 scenario problem). In both biological and generic systems, manylp dominates when scenarios share critical regions (left of the dotted 512 pool-capacity threshold) but incurs overhead when highly incoherent scenarios mandate unique simplex solves.

## TABLES

| Solver | Platform | Native batching | Per-LP certification | Steady-state throughput (LP/s) | Max relative objective error |
|---|---|---|---|---|---|
| manylp (plain FBA) | CPU | Yes | Yes | 47,889 | 0 (reference) |
| manylp (pFBA-unique) | GPU | Yes | Yes | 22,827 | $2.6 \times 10^{-11}$ |
| HiGHS (dual simplex) | CPU | No | No (status flag) | 3,986 | $3.3 \times 10^{-11}$ |
| Gurobi 13 (dual simplex) | CPU | No | No (status flag) | 2,958 | $2.1 \times 10^{-9}$ |
| FICO Xpress (dual simplex) | CPU | No | No (status flag) | 2,712 | $2.1 \times 10^{-9}$ |
| GLPK | CPU | No | No (status flag) | 1,835 | $3.4 \times 10^{-10}$ |
| MPAX (batched PDHG) | GPU | Yes | No (ε-optimal) | 2.50 | $2.5 \times 10^{-2}$ (12% solved) |
| NVIDIA cuOpt (PDLP) | GPU | Yes | No (ε-optimal) | 0.40 | $1.0 \times 10^{-6}$ |
| NVIDIA cuOpt (PDLP + crossover) | GPU | Yes | No (status flag) | 0.21 | $1.1 \times 10^{-11}$ |

**Table 1:** Speed, accuracy, and capability of evaluated LP solvers on a coherent genome-scale FBA workload. Performance metrics are recorded over 185,088 LPs derived from consecutive dFBA time steps. Throughput represents steady-state evaluations per second. Exact solvers verify the optimum, while first-order GPU solvers (cuOpt PDLP, MPAX) struggle with the degenerate biological matrices, requiring crossover to achieve exactness. manylp uniquely provides rigorous per-LP mathematical certification for optimality, uniqueness, and infeasibility.

| Biological quantity | Minimum | pFBA-unique | Maximum | Variation (max/min) |
|---|---|---|---|---|
| Acetate (mM) | 0.209 | 0.436 | 0.883 | 4.2x |
| Butyrate (mM) | 0.115 | 0.172 | 0.343 | 3.0x |
| H2 (mM) | 1.17 | 1.25 | 1.79 | 1.5x |
| Propionate (mM) | $4.4 \times 10^{-6}$ | $7.0 \times 10^{-5}$ | $1.6 \times 10^{-3}$ | 364x |
| B. fragilis biomass (gDW $L^{-1}$) | $6.57 \times 10^{-3}$ | – | $9.38 \times 10^{-3}$ | 1.43x |
| B. thetaiotaomicron biomass (gDW $L^{-1}$) | $1.80 \times 10^{-2}$ | – | $2.13 \times 10^{-2}$ | 1.18x |
| B. fragilis relative abundance | 15.7% | – | 20.6% | 1.31x |

**Table 2:** Extreme bounds of biological predictions across 49 equally optimal flux policies. Final trajectory values at 48 hours for a 12-species gut community model. Min and max values span the entirety of the alternative-optima envelope, demonstrating the severe variability hidden within standard FBA vertex solutions.

| Simulation configuration | Total LPs | Original backend time (s) | manylp GPU time (s) | Speedup | Total simplex solves (manylp) |
|---|---|---|---|---|---|
| Ensemble (E=1) | 5,772 | 653.6 | 7.7 | 84x | 27 |
| Ensemble (E=16) | 92,352 | 10,920 | 17.2 | 635x | 72 |
| Ensemble (E=256) | 1,477,632 | 167,318 | 188.7 | ~890x | 72 |
| Ensemble (E=1,024) | 5,910,528 | 669,273 | 776.1 | ~860x | 72 |
| Spatial grid (8x8) | 7,680 | 868.5 | 8.7 | ~100x | – |
| Spatial grid (16x16) | 30,720 | 3,474 | 16.9 | ~206x | – |
| Spatial grid (32x32) | 122,880 | 13,897 | 50.8 | ~274x | – |

**Table 3:** End-to-end acceleration of dynamic community and spatial simulations in μODE. All runtimes include the simulator's full kinetic evaluation, ODE integration, and bookkeeping overhead. Ensemble simulations model a 12-species gut community for 48 hours (dt=0.1h). Spatial simulations model 1 hour of diffusion across 2D grids.

| Biological scenario | Integration scheme | Total LPs | Wall time (s) | Max relative biomass error | Max relative product error |
|---|---|---|---|---|---|
| Gut community (48 h) | Euler, dt=0.1h | 5,772 | 6.0 | $4.5\times10^{-3}$ | $1.5\times10^{-2}$ |
| | Euler, dt=0.05h | 11,532 | 9.1 | $2.2\times10^{-3}$ | $7.1\times10^{-3}$ |
| | Euler, dt=0.01h | 57,612 | 33.4 | $4.0\times10^{-4}$ | $1.3\times10^{-3}$ |
| | RK45, rtol $10^{-4}$ | 15,864 | 14.4 | $5.4\times10^{-5}$ | $1.4\times10^{-3}$ |
| | RK45, rtol $10^{-6}$ | 20,616 | 14.6 | $4.4\times10^{-5}$ | $3.8\times10^{-4}$ |
| *E. coli* diauxie (12 h) | Euler, dt=0.1h | 242 | 0.13 | $7.5\times10^{-2}$ | 6.5 |
| | Euler, dt=0.01h | 2,402 | 0.92 | $7.9\times10^{-3}$ | 0.21 |
| | RK45, rtol $10^{-6}$ | 1,408 | 0.77 | $8.1\times10^{-5}$ | $1.9\times10^{-3}$ |
| | RK45, rtol $10^{-6}$ + regime events | 2,586 | 1.96 | $8.1\times10^{-5}$ | $2.0\times10^{-3}$ |

**Table 4:** Integration accuracy and computational cost across fixed and adaptive schemes. Maximum relative errors for biomass and metabolic products (acetate, butyrate, and H2 for the gut model); glucose and acetate for the diauxie model) are calculated against highly refined Euler reference trajectories. The RK45 adaptive schemes leverage manylp's microsecond evaluation speeds to deliver multiple orders of magnitude of error reduction.

| Platform | Ensemble size (E) | Strategy | Total solve time (s) | Simplex solves | Repair time (s) | Certify time (s) |
|---|---|---|---|---|---|---|
| CPU | 64 | Eager | 20.4 | 864 | 13.2 | 4.82 |
| CPU | 64 | Progressive | 11.0 | 76 | 3.60 | 4.91 |
| CPU | 64 | Progressive + atlas | 8.36 | 0 | 0.02 | 5.49 |
| GPU | 1,024 | Eager | 37.8 | 864 | 17.1 | 8.16 |
| GPU | 1,024 | Progressive | 23.9 | 76 | 3.76 | 7.30 |
| GPU | 1,024 | Progressive + atlas | 21.1 | 0 | 0.03 | 7.50 |

**Table 5:** Cold-start ablation and the impact of progressive repair and basis atlases. Performance metrics for resolving initial cache misses on a 12-species gut community model simulated for 48 hours. Eager repair solves all uncertified members, while progressive repair uses farthest-point sampling to solve representative LPs. The progressive + atlas variant utilizes a pre-computed basis atlas trained on a distinct diet ensemble, entirely eliminating the need for new simplex solves on the unseen evaluation ensemble.

| Workload | manylp metric | Best baseline metric | Baseline method |
|---|---|---|---|
| Coherent FBA (185,088 LPs) | 47,889 LP/s | 7,058 LP/s | Per-LP bunching (8 proc.) |
| dFBA ensemble (E=1,024) | 140,527 LP/s | 20,903 LP/s | Per-LP bunching (32 proc.) |
| *E. coli* diauxie (E=32) | 0.41 s / trajectory | 1.9 s / trajectory | dfba (event method) |
| iJO1366 (E=8, single core, rtol $10^{-8}$) | 32.7 s / trajectory | 25.6 s / trajectory | dfba (fixed interval) |
| iJO1366 (Event method, 0.5 h) | Stable integration | 229-495 s<br>(20% biomass error) | dfba (event method) |

**Table 6:** Performance baseline comparison against classical bunching and trajectory-based event simulators. Throughput and per-trajectory timings evaluate the limits of structural reuse methodologies. While classical bunching captures reuse efficiently for single-threaded plain FBA, it fails to match manylp's parallelized dense matrix throughput. Trajectory simulators minimize LP calls but cannot compete with manylp's microsecond evaluation speed during ensemble processing.

## METHODS

### Problem formulation and the bounds-parametric class

Dynamic simulators generate large families of closely related linear programs. manylp organizes these structurally identical problems into a group where the stoichiometric matrix A (dimensions m×n) and the ordered list of biological objectives $c_1,\ldots,c_K$ remain constant. Individual members within a batch supply specific upper and lower bounds. To standardize the problem, structural constraints are converted into unified row logicals s=Ax, isolating all member-specific variation strictly into the variable bounds. The problem is evaluated in a normalized z-space:

Equation 1. $$\bar{A}z = 0, \quad \bar{A} = [A, -I], \quad z = (x, s), \quad l^b \leq z \leq u^b$$

This transformation ensures the core matrix Ā retains full row rank m. A basis B is defined by selecting m linearly independent columns of Ā and assigning each nonbasic variable to its lower bound, upper bound, or a free state at zero [21].

### Mathematical certification conditions

The certify-and-repair architecture relies on validating a cached basis against new environmental bounds. When bounds shift, the values of the basic variables $z_B$ adjust as an affine function of the nonbasic variables $z_N$, anchored at a template bound state $\hat{z}_N$:

Equation 2. $$z_B = h + T\left(z_N - \hat{z}_N\right)$$

where $T=-B^{-1}\bar{A}_N$, and h is the anchor point. Conversely, the lexicographic reduced costs $d_k$ for stage k depend entirely on the matrix and objectives, completely unaffected by the shifting bounds [60]:

Equation 3. $$d_k = c_k - \bar{A}^{T} B^{-T} c_{k,B}$$

manylp certifies lexicographic optimality by computationally verifying two conditions per LP in double-precision arithmetic:

1. <u>Primal feasibility:</u> The resulting basic variables must strictly respect the member bounds: $l_B \le z_B \le u_B$.
2. <u>Lexicographic dual feasibility:</u> For every nonbasic variable j, the first non-zero entry in the vector $(d_{1j},\ldots,d_{Kj})$ must possess a sign matching its bound status (e.g., ≤0 at the lower bound, ≥0 at the upper bound) [52,74].

**Uniqueness condition –** The solution is verified as geometrically unique if, for every nonbasic variable j not fixed at a bound ($l_j<z_j<u_j$), the lexicographic reduced-cost vector contains at least one non-zero entry.

**Farkas infeasibility test –** For biological states that trigger infeasibility (e.g., severe nutrient starvation), the solver caches Farkas vectors [69] y. For $r=\bar{A}^{T}y$, any feasible z requires $r^{T}z=0$. A state is mathematically certified as infeasible if the maximum possible value of $r^{T}z$ over the bound box is strictly negative:

Equation 4. $$max\left(r^{T}z\right) < - \tau$$

where the tolerance τ is not a static epsilon, but a dynamic rounding-error bound of the test itself (with a relative floor of $10^{-11}$) [85].

For additional mathematical details on lazy sparse matrix updates, active nonbasic column identification, and dynamic double-precision error tolerances, see Supplementary Material (<u>Lazy parameter-sparse affine laws and dynamic certification bounds</u>).

### Batched hardware acceleration and lazy evaluation

To process massive batches efficiently across hardware accelerators, manylp avoids calculating the full affine transformation T for every reaction. It evaluates a parameter-sparse, lazy affine law, computing only the columns of T corresponding to parametric variables that actively deviate from the template. Because microbial organisms are typically constrained by only a handful of limiting nutrients simultaneously, only a small fraction of exchange bounds activate per basis.

By isolating these active columns, checking an entire batch of Bg members against a basis is reduced to a single dense matrix product. On a GPU, this product executes via cuBLAS, enclosed by custom CUDA kernels that fuse the element-wise primal feasibility checks and trigger atomic operations only upon mathematical failure. For CPU-bound execution, equivalent Numba kernels run concurrently across members alongside BLAS [86].

Cached bases are maintained in a dynamic pool (default capacity of 512 bases) ordered by recent certification activity so that the most frequently visited metabolic regimes are tested first. When a new basis enters the pool, its sparse LU factorization is retained on the host while its active transformation columns are materialized lazily on the target device as new environmental bounds activate them. Because optimal bases and their associated critical regions depend strictly on the stoichiometry and objective vectors rather than the transient bounds, manylp supports serializing this pool to disk as a persistent basis atlas. When loaded into subsequent simulations of the same organism or community under unseen environmental perturbations, the pre-populated atlas allows the solver to immediately certify known metabolic regimes via matrix multiplication, bypassing cold-start simplex exploration entirely.

**Progressive repair and yield guarding**

When new biological states fall outside the regions covered by the currently cached bases, manylp isolates the uncertified members and identifies a representative subset to solve. To ensure diverse geometric coverage of the bounding space, these representatives are selected using farthest-point sampling in normalized bound space. The selected linear programs are then routed to a lexicographic dual simplex algorithm (HiGHS), which is warm-started from the most mathematically proximate cached basis. To properly handle biological degeneracy, the simplex carves out the exact optimal face by sequentially fixing nonbasic variables with non-zero reduced costs at their bounds between optimization stages [7,23,74].

To handle numerical instability gracefully, the solver utilizes a fallback retry ladder, escalating from warm starts to cold starts, and finally to cold starts enhanced with advanced presolve mechanics [22]. Once a new basis is successfully verified, it is immediately propagated against all remaining uncertified members in the batch. To prevent performance degradation on highly incoherent batches where caching fails, a yield guard continuously monitors propagation efficiency. If a repair round resolves fewer than 1.5 members per newly discovered basis, the batched propagation is halted, and remaining members are solved individually in a direct mode.

This process is summarized in Algorithm 1 below.

```
Algorithm 1: Batched Certify-and-Repair
Input: Batch of LPs with varying bounds, Pool of cached bases B, Farkas cache F
Output: Certified vertex solutions for all LPs

while uncertified LPs remain in Batch do
    // Phase 1: matrix certification
    Evaluate z_B = h + T(z_N - ẑ_N) for all uncertified LPs across bases in B
    Mark LPs satisfying bounds (l ≤ z_B ≤ u) as certified optimal

    // Phase 2: Farkas infeasibility check
    Evaluate max(r^T z) for remaining LPs across vectors in F
    Mark LPs where max(r^T z) < -τ as certified infeasible

    // Phase 3: progressive repair
    Sample representative LPs from the remaining uncertified batch
    for each representative LP do
        Solve via warm-started lexicographic dual simplex
        Verify new basis internal status in double-precision arithmetic
        Add verified basis to pool B
    end for

    // Phase 4: yield guarding
```

```
    if newly discovered bases certify < 1.5 LPs on average then
        Switch remaining LPs to direct mode (solve independently)
    end if
end while
```

Rather than solving uncertified LPs in an arbitrary order, farthest-point sampling selects representative problems whose normalized bound vectors are maximally separated from one another and from previously sampled points. In a biological ensemble or spatial grid, this ensures that the repair phase probes distinct environmental extremes (such as different nutrient-depletion regimes) simultaneously rather than redundantly solving neighboring trajectories that share the same critical region. Each selected representative is then warm-started using the cached basis whose template bounds are closest in bound space, minimizing the number of dual simplex pivots required to reach the new optimal face.

For the explicit normalized coordinate equations and in-place lexicographic face-carving mechanics, see Supplementary Material (Farthest-point sampling and lexicographic face carving).

**Enforcing flux uniqueness and constructing envelopes**

Standard FBA leaves internal metabolic fluxes undetermined. To ensure reproducibility, we adopt a lexicographic flux rule: maximize primary biological growth, then minimize the total absolute sum of all fluxes, operating as standard parsimonious FBA (pFBA), and finally maximize a generic, fixed direction vector drawn from a seeded pseudo-random stream to break remaining ties [7,50,53]. During the parsimonious stage, reversible reactions are split into strictly non-negative components to maintain structural independence from the variable bounds. This precise sequence mathematically guarantees convergence to a single, certified vertex. We call this rule pFBA-unique. manylp certifies, per LP, that the resulting optimum is unique.

To move beyond single trajectories and quantify biological uncertainty, manylp generates alternative-optima envelopes. Rather than replacing the third uniqueness stage, this is achieved by inserting diverse, targeted selection policies, such as maximizing or minimizing the specific secretion of various fermentation products, before the parsimonious $L_1$ minimization stage [36]. This isolates the targeted biological objective while ensuring the subsequent stages still collapse any remaining degeneracy, integrating dozens of extreme boundary trajectories simultaneously to yield the complete envelope of dynamic predictions the biological model permits.

In our human gut microbiome experiments, the 49-policy alternative-optima envelope was constructed by combining the baseline three-stage pFBA-unique policy with 48 four-stage probe policies evaluated in lockstep. Specifically, 16 extremal policies inserted a secondary objective that either maximized or minimized the flux of one of eight major fermentation products (acetate, butyrate, propionate, H2, and four additional cross-feeding byproducts) immediately after primary biomass maximization, followed by parsimonious L1 minimization and random tie-breaking. The remaining 32 policies replaced the metabolite-specific stage with 32 seeded random unit objective vectors to sample the interior boundaries of the optimal growth face before collapsing residual degeneracy.

For the complete mathematical formulation of the split-variable parsimonious hierarchy and the precise composition of the 48 probe policies, see Supplementary Material (Formulation of pFBA-unique and alternative-optima envelopes).

**Dynamic integration schemes and benchmark workloads**

To evaluate solver performance across realistic and stress-test regimes, we constructed four biological benchmark suites alongside a generic non-biological benchmark. First, a temporally coherent workload (185,088 LPs) was recorded sequentially across the first 24 hours of a 12-species human gut community model reconstructed via gapseq and simulated on a Western diet in µODE. Second, a weakly coherent snapshot workload (23,808 LPs) was extracted by subsampling the trajectory every 1.6 simulated hours (every 16th integration step) to strip away step-to-step temporal locality. Third, diet-perturbed community ensembles (scaling from E=1 to 16,384 members over 48 hours) were generated by applying independent log-normal perturbations to the dietary uptake bounds of each ensemble member. Fourth, 2D spatial reaction-diffusion grids (8x8, 16x16, and 32x32 cells simulated for 1 hour) were executed in µODE's COMETS-style spatial engine by batching all grid cells for a given species into a single manylp evaluation at each time step. Finally, to evaluate generalization outside systems biology, 44 standard problems from the Netlib LP repository were expanded into 1,024-member batches by independently perturbing their right-hand-side bounds.

For time integration benchmarks on the 12-species gut community (48 hours) and the classic *E. coli* core diauxic shift (12 hours of aerobic growth transitioning from glucose to acetate), we compared standard fixed-step explicit Euler integration (dt=0.1 h, 0.05 h, and 0.01 h) against an adaptive Dormand-Prince embedded Runge-Kutta scheme (RK45, evaluated at relative tolerances of $10^{-4}$, $10^{-6}$, and $10^{-8}$) against ultra-fine reference Euler trajectories (dt=$10^{-3}$ h for the community and $10^{-4}$ h for diauxie). In the regime-event splitting variant, the integrator additionally tracked continuous zero-crossings of the basic variables against their bounds to locate exact basis-switching timestamps during a step.

For the complete ODE mass-balance specifications, batched RK45 stage structures, and zero-crossing event root-finding algorithms, see Supplementary Material (Coupled dFBA integration and regime-event location).

**Implementation stack and audit protocol**

The manylp software architecture is built in Python, heavily leveraging NumPy and SciPy for core operations, CuPy for GPU arrays, and Numba for fused CPU kernel acceleration [86–88]. Dynamic integration is handled through SciPy's embedded Runge-Kutta (Dormand-Prince RK45) wrappers [76], which execute efficiently over the smooth, affine critical regions.

Empirical benchmarking was executed on a shared computational node utilizing an NVIDIA A100-PCIE-40GB GPU alongside four Intel Xeon Platinum 8276L CPUs. General-purpose per-LP baseline solvers, including HiGHS, Gurobi, FICO Xpress, GLPK, and OR-Tools GLOP, were allocated 32 parallel worker processes to maximize their throughput. First-order GPU baselines, including NVIDIA cuOpt and MPAX, were restricted to relative and absolute tolerance of $10^{-6}$ [27].

To guarantee the validity of the findings, every single flux vector generated by any solver was evaluated by a unified, independent audit harness. This harness recomputed primal infeasibility, objective error, status agreement, and exchange-flux derivations against an exact reference. The reference itself was cross-validated on a 400-LP subset using an independent sequential HiGHS implementation chained through objective-bound rows [23,74], verifying absolute mathematical agreement to within a tolerance of $1.7\times10^{-11}$.

For comprehensive parameter configurations across all 25 solver packages, custom baseline implementations, and exact audit error equations, see Supplementary Material (Baseline solver configurations and independent audit metrics).

## CODE AVAILABILITY

manylp is open-source software released under the MIT license at https://github.com/cumbof/manylp. It is written in Python (version 3.10 or later) and depends on NumPy, SciPy, and the HiGHS solver (highspy). GPU execution additionally requires CuPy and a CUDA 12 device, and the fused multi-threaded CPU kernels require Numba. The repository includes a test suite that checks manylp's certified answers against an independent reference solver on both CPU and GPU. All results in this manuscript can be reproduced from the same repository.

The µODE simulator, used for the community and spatial simulations, is also open-source software released under the MIT license at https://github.com/cumbof/muODE.

**Lazy parameter-sparse affine laws and dynamic certification bounds**

For a metabolic network with stoichiometric matrix A (dimensions m×n) and K ordered linear objectives $c_1$, …, $c_K$, manylp converts structural constraints into row logicals s=Ax and operates in the unified variable space z=(x, s) of dimension n+m with the augmented matrix Ā=[A, −I]. Across a batch of Bg linear programs, each problem instance b is completely specified by its lower and upper bound vectors (lb, ub).

A cached basis partitions the n+m columns of Ā into m basic variables (forming a nonsingular m×m basis matrix B) and n nonbasic variables fixed at their lower bound, their upper bound, or zero (for free variables). For any LP in the batch, the nonbasic vector zN is assembled directly by reading the corresponding entries from that member's lower or upper bound vectors.

When a new basis B is discovered from a template LP with nonbasic state ẑN, the solver stores the sparse LU factorization of B on the host CPU along with the template basic anchor vector $h = -B^{-1}$ ĀN ẑN. Rather than computing the full m×n sensitivity matrix $T = -B^{-1}$ ĀN upfront, manylp identifies only the active nonbasic columns that actually change across the current batch (where |zN − ẑN| > 0). For any newly activated nonbasic variable not yet present in the basis cache, its sensitivity column in T is computed on demand via a single sparse triangular solve against the stored LU factors of B and appended to the dense device-resident submatrix T_act. Evaluating the basic variables zB across all Bg batch members then reduces to a single dense matrix multiplication: zB = h + T_act (zN − ẑN), restricted solely to the active columns.

To protect against floating-point accumulation errors in double-precision arithmetic (machine epsilon $\varepsilon \approx 2.22\times10^{-16}$), certification checks apply dynamic, magnitude-scaled tolerances per LP:

- <u>Primal feasibility:</u> A batch member is certified primal feasible under a cached basis if every basic variable satisfies lB − τP ≤ zB ≤ uB + τP, where the component-wise tolerance τP scales with the variable and bound magnitudes as $\tau P = 10^{-9} \times \max(1, |zB|, |lB|, |uB|)$.
- <u>Lexicographic dual feasibility and uniqueness:</u> For each optimization stage k, the reduced-cost vector $dk = ck - \bar{A}^{T} B^{-T} ck,B$ depends only on the network matrix and objectives and is computed once when the basis is cached. Any reduced-cost entry with magnitude $|dk,j| \le 10^{-9} \times (1 + \max |ck|)$ is treated as numerically zero. A basis is certified dual feasible if, for every nonbasic variable j, the first numerically non-zero value in its lexicographic tuple (d1j, …, dKj) is ≤ 0 when fixed at its lower bound, ≥ 0 when fixed at its upper bound, and zero across all stages if free. The optimal flux vector for a member is certified as geometrically unique if every nonbasic variable with a non-fixed bound interval (lj < uj) has at least one numerically non-zero entry in (d1j, …, dKj).
- <u>Farkas infeasibility certificate:</u> When the simplex repair phase encounters an infeasible LP, a dual Farkas ray y is extracted and mapped to $r = \bar{A}^{T} y$ (normalized so that max |rj| = 1). Because any feasible solution z satisfies Āz = 0 and therefore $r^{T}z = 0$, the maximum possible value of $r^{T}z$ over a member's bound box [l, u] can be evaluated directly without solving an LP by summing rj × uj for all positive entries (rj > 0) and rj × lj for all negative entries (rj < 0). A member is certified infeasible if this sum is strictly less than −τF, where the dynamic rounding-error bound τF is set to $\max(10^{-11}, 10^{3} \times \varepsilon \times \mathrm{sum}(|rj| \times \max(|lj|, |uj|)))$.

**Farthest-point sampling and lexicographic face carving**

When a subset of LPs in a batch remains uncertified after testing all cached bases and Farkas rays, manylp selects representative problems for simplex repair using Gonzalez's farthest-point sampling in normalized bound space. First, each varying bound coordinate across the uncertified LPs is linearly rescaled to the [0, 1] interval using (bound − min_bound) / (max_bound − min_bound + $10^{-12}$). The first representative LP is chosen as the uncertified member with the smallest maximum primal bound violation against the cached basis pool, and subsequent representatives are selected iteratively by picking the uncertified member with the largest Euclidean distance to the already-sampled set. Each selected representative is warm-started using the cached basis whose template bounds are closest in Euclidean distance.

To keep the row dimension m constant across all K lexicographic optimization stages, manylp avoids appending objective-constraint rows ($c_k^T x \geq z_k^*$) to the stoichiometric matrix during basis repair. Instead, after stage k is solved to optimality via dual simplex, the optimal face is carved out in place by locking every nonbasic variable with a non-zero reduced cost to its active bound (setting $l_j = u_j$) before solving stage k+1. Once the final stage K converges, the original member bounds are restored and the terminal basis is verified in double-precision arithmetic and added to the basis pool.

### Formulation of pFBA-unique and alternative-optima envelopes

To express the parsimonious L1 flux minimization objective as a purely linear function that does not change when variable bounds shift, reversible reactions are split into non-negative forward and backward components ($x = x+ - x-$, with $x+, x- \geq 0$) during model initialization. The three-stage pFBA-unique hierarchy then executes sequentially on the same basis structure:

1. Stage 1 (Primary biological objective): Maximize primary cellular growth (biomass production).
2. Stage 2 (Parsimonious L1 minimization): Minimize the sum of all split internal and exchange fluxes ($x+ + x-$) over the Stage 1 optimal face.
3. Stage 3 (Deterministic tie-breaking): Maximize a fixed, dense objective vector drawn from a seeded pseudo-random stream over the Stage 2 optimal face to isolate a unique vertex.

To construct the 49-policy alternative-optima envelope, 48 four-stage probe policies are evaluated in lockstep alongside the baseline three-stage pFBA-unique policy. Each four-stage policy inserts a probe objective at Stage 2 immediately after primary biomass maximization, shifting parsimonious L1 minimization to Stage 3 and random tie-breaking to Stage 4. The 48 probe objectives consist of 16 extremal metabolite policies (maximizing and minimizing exchange fluxes for eight fermentation products, including acetate, butyrate, propionate, and $H_2$) and 32 seeded random unit objective vectors that sample alternative directions across the optimal growth face.

### Coupled dFBA integration and regime-event location

In community dFBA simulations within μODE, extracellular metabolite concentrations C(t) and microbial species biomasses X(t) evolve over time for each ensemble member according to standard mass-balance differential equations: the biomass of species i grows at rate $dX_i/dt = \mu_i(t) \times X_i(t)$, and the concentration of extracellular metabolite m changes at rate $dC_m/dt = \text{sum}_i (v_{i,m}(t) \times X_i(t))$, where $\mu_i(t)$ and $v_{i,m}(t)$ are the certified growth rate and exchange flux returned by manylp under concentration-dependent uptake bounds. During adaptive Dormand-Prince RK45 integration, all ensemble members (or all 49 envelope policies) are grouped into a single batched manylp call at each internal Runge-Kutta stage evaluation. In the regime-event splitting configuration (Table 4), the integrator additionally monitors continuous zero-crossing functions measuring the distance of basic variables to their lower and upper bounds ($z_B(t) - l_B(t) = 0$ and $u_B(t) - z_B(t) = 0$) to pinpoint exact basis-switching timestamps via root-finding.

### Baseline solver configurations and independent audit metrics

General-purpose CPU baselines (HiGHS, Gurobi 13, FICO Xpress, GLPK, OR-Tools GLOP, SciPy linprog, and OSQP) were benchmarked using 32 parallel worker processes (p32), with warm-started modes reusing the optimal basis from the preceding time step along each trajectory. For the classical bunching baselines (Table 6), we implemented the Haugland-Wallace procedure across 8 and 32 worker processes, storing sparse LU factorizations of optimal bases and testing primal feasibility via per-LP forward and backward triangular solves before falling back to warm-started dual simplex. First-order GPU baselines (NVIDIA cuOpt PDLP, MPAX, and our custom batched primal-dual hybrid gradient solver ourpdhg) were executed on the NVIDIA A100-PCIE-40GB GPU with relative and absolute tolerances set to $10^{-6}$ (and $10^{-4}$ for the snapshot sensitivity test in Table S2) under a 15-minute wall-clock timeout per workload.

Every flux vector x returned by any solver configuration was verified offline by an independent double-precision audit harness against the cross-validated reference solution $x_{ref}$ (with optimal objective $z_{ref} = c1^T x_{ref}$) using three primary accuracy metrics (reported in Tables 1, S1, S2, and S3):

- Maximum objective error (Max obj. error): The maximum across all LPs of the relative objective discrepancy $|c1^T x - z_{ref}| / \max(1, |z_{ref}|)$.
- Maximum primal infeasibility (Max infeasibility): The worst-case violation across all LPs of either the steady-state mass balance or the variable bounds, computed as $\max(|Ax - s|, \max(0, l - z), \max(0, z - u))$.
- Exchange flux deviation (Exchange dev. median / max): The median and maximum across all solved LPs of the largest absolute exchange flux difference $\max |x_{ex} - x_{ref,ex}|$ over all extracellular exchange reactions.

## SUPPLEMENTARY TABLES

| Configuration | Device | LP/s (overall) | LP/s (steady) | Done | Solved | Status agreement | Max obj. error | Max infeasibility | Exchange dev. median / max |
|---|---|---|---|---|---|---|---|---|---|
| **manylp-cpu-fba** | **CPU** | **17,916** | **47,889** | **100%** | **100%** | **100%** | **0** | **1.0e-10** | **0.71/1.6** |
| **manylp-gpu-fba** | **GPU** | **17,762** | **31,844** | **100%** | **100%** | **100%** | **0** | **1.0e-10** | **0.71/1.6** |
| **manylp-cpu-pfba-unique** | **CPU** | **15,569** | **24,206** | **100%** | **100%** | **100%** | **2.6e-11** | **4.2e-12** | **0/0** |
| **manylp-gpu-pfba-unique** | **GPU** | **14,821** | **22,827** | **100%** | **100%** | **100%** | **2.6e-11** | **4.2e-12** | **0/0** |
| highs-simplex-warm-p32 | CPU | 3,866 | 3,986 | 100% | 100% | 100% | 3.3e-11 | 2.6e-11 | 0.85/2.7 |
| gurobi-dual-warm-p32 | CPU | 2,926 | 2,958 | 100% | 100% | 100% | 2.1e-9 | 6.1e-7 | 0.89/1.8 |
| xpress-dual-p32 | CPU | 2,694 | 2,712 | 100% | 100% | 100% | 2.1e-9 | 6.1e-7 | 0.86/1.7 |
| glpk-p32 | CPU | 1,771 | 1,835 | 100% | 100% | 100% | 3.4e-10 | 2.6e-7 | 0.70/1.6 |
| glop-p32 | CPU | 1,442 | 1,465 | 100% | 100% | 100% | 1.2e-8 | 2.0e-7 | 0.71/2.5 |
| highs-lex-pfba-unique-p32 | CPU | 745 | 816 | 100% | 100% | 100% | 2.6e-11 | 2.0e-11 | 1.6e-13/6.7e-12 |
| **manylp-gpu-perlp-fba** | **GPU** | **726** | **781** | **100%** | **100%** | **100%** | **4.3e-11** | **1.0e-10** | **0.71/1.6** |
| scipy-p32 | CPU | 584 | 582 | 100% | 100% | 100% | 3.3e-11 | 2.5e-11 | 0.89/2.7 |
| highs-ipm-cold-p32 | CPU | 434 | 438 | 100% | 100% | 100% | 3.3e-11 | 1.6e-10 | 0.85/2.7 |
| ourpdhg-e1e-6 | GPU | 63 | 65 | 30.7% | 0% | 0% | 0.71 | 0.23 | 3.5/26 |
| cuopt-concurrent-e1e-6 | GPU + CPU | 17 | 17 | 8.2% | 100% | 100% | 1.2e-9 | 9.0e-8 | 0.83/2.5 |
| mpax-e1e-6 | GPU | 1.79 | 2.46 | 0.9% | 12.3% | 12.3% | 2.5e-2 | 6.4e-3 | 0.63/2.2 |
| cuopt-pdlp-batch-e1e-6 | GPU | 0.48 | 0.40 | 0.2% | 100% | 100% | 1.0e-6 | 1.4e-6 | 0.09/1.8 |
| cuopt-barrier-xover-e1e-6 | GPU | 9.65 | 9.63 | 4.7% | 0.01% | 0.01% | 4.4e-2 | 8.5e-4 | 0.05/3.4 |
| cuopt-pdlp-xover-e1e-6 | GPU | 0.27 | 0.21 | 0.2% | 100% | 100% | 1.1e-11 | 9.0e-8 | 0.16/2.0 |

**Table S1:** Complete solver performance on the temporally coherent dFBA workload (185,088 LPs).

| Configuration | Device | LP/s (overall) | LP/s (steady) | Done | Solved | Status agreement | Max obj. error | Max infeasibility | Exchange dev. median / max |
|---|---|---|---|---|---|---|---|---|---|
| **manylp-gpu-fba** | **GPU** | **4,930** | **–** | **100%** | **100%** | **100%** | **2.7e-11** | **1.0e-10** | **0.73/1.6** |
| **manylp-cpu-pfba-unique** | **CPU** | **4,860** | **–** | **100%** | **100%** | **100%** | **1.6e-11** | **4.2e-12** | **0/1.1e-11** |
| **manylp-gpu-pfba-unique** | **GPU** | **4,252** | **–** | **100%** | **100%** | **100%** | **1.6e-11** | **4.2e-12** | **0/1.1e-11** |
| highs-simplex-warm-p32 | CPU | 3,491 | – | 100% | 100% | 100% | 1.6e-11 | 1.3e-11 | 0.83/2.7 |
| **manylp-cpu-fba** | **CPU** | **3,477** | **–** | **100%** | **100%** | **100%** | **2.7e-11** | **1.0e-10** | **0.73/1.6** |
| gurobi-dual-warm-p32 | CPU | 2,594 | 3,032 | 100% | 100% | 100% | 7.6e-10 | 2.2e-7 | 0.80/1.8 |
| xpress-dual-p32 | CPU | 2,326 | 2,811 | 100% | 100% | 100% | 7.6e-10 | 2.6e-7 | 0.87/2.1 |
| glpk-p32 | CPU | 1,303 | 1,818 | 100% | 100% | 100% | 3.4e-10 | 2.6e-7 | 0.72/1.7 |
| glop-p32 | CPU | 1,271 | 1,306 | 100% | 100% | 100% | 4.7e-9 | 2.2e-8 | 0.70/2.0 |
| highs-lex-pfba-unique-p32 | CPU | 731 | 869 | 100% | 100% | 100% | 1.6e-11 | 7.7e-12 | 1.3e-13/9.5e-12 |
| **manylp-gpu-perlp-fba** | **GPU** | **687** | **–** | **100%** | **100%** | **100%** | **9.9e-12** | **1.0e-10** | **0.72/1.6** |
| highs-simplex-cold-p32 | CPU | 622 | – | 100% | 100% | 100% | 1.6e-11 | 1.4e-11 | 0.89/2.7 |
| scipy-p32 | CPU | 512 | 524 | 100% | 100% | 100% | 1.6e-11 | 1.4e-11 | 0.89/2.7 |
| highs-ipm-cold-p32 | CPU | 449 | – | 100% | 100% | 100% | 1.6e-11 | 9.4e-11 | 0.85/2.7 |
| gurobi-barrier-cold-p32 | CPU | 142 | 142 | 100% | 100% | 100% | 1.6e-8 | 1.0e-6 | 0.57/2.6 |
| ourpdhg-e1e-6 | GPU | 61 | – | 100% | 0% | 0% | 0.71 | 0.26 | 3.2/26 |
| highs-pdlp-cold-p32 | CPU | 51 | – | 100% | 0% | 0% | 4.5e-2 | 0 | 0.04/3.5 |
| cuopt-concurrent-e1e-6 | GPU + CPU | 17 | – | 64.0% | 100% | 100% | 5.2e-7 | 2.0e-7 | 0.83/2.6 |
| cuopt-barrier-xover-e1e-6 | GPU | 10 | 10 | 38.4% | 0.04% | 0.04% | 4.5e-2 | 4.0e-3 | 0.03/3.5 |
| cuopt-barrier-e1e-6 | GPU | 10 | – | 38.2% | 0.06% | 0.06% | 4.5e-2 | 6.3e-5 | 0.03/3.5 |
| cuopt-dual-simplex-e1e-6 | CPU | 7.10 | 7.11 | 26.9% | 100% | 100% | 9.6e-10 | 2.6e-7 | 0.83/2.6 |
| osqp-p32 | CPU | 3.33 | 3.56 | 12.6% | 18.1% | 18.1% | 4.5e-2 | 2.3e-6 | 0.54/3.4 |
| mpax-e1e-6 | GPU | 1.90 | – | 7.3% | 11.7% | 11.7% | 4.4e-2 | 1.3e-2 | 0.64/2.1 |
| mpax-e1e-4 | GPU | 1.88 | – | 7.3% | 22.1% | 22.1% | 4.4e-2 | 1.2e-2 | 0.63/2.1 |
| ortools-pdlp-p32 | CPU | 1.66 | 1.50 | 6.5% | 95.8% | 95.8% | 2.0e-5 | 9.6e-7 | 0.60/2.1 |
| cuopt-pdlp-batch-e1e-6 | GPU | 0.48 | – | 1.9% | 100% | 100% | 1.0e-6 | 1.4e-6 | 0.09/1.8 |
| cuopt-pdlp-e1e-4 | GPU | 0.34 | 0.27 | 1.9% | 100% | 100% | 1.0e-4 | 5.3e-4 | 0.01/1.8 |
| cuopt-pdlp-warm-e1e-6 | GPU | 0.27 | – | 1.9% | 100% | 100% | 1.0e-6 | 1.4e-6 | 0.09/1.8 |
| cuopt-pdlp-e1e-6 | GPU | 0.27 | – | 1.9% | 100% | 100% | 1.0e-6 | 1.4e-6 | 0.09/1.8 |
| cuopt-pdlp-xover-e1e-6 | GPU | 0.27 | 0.21 | 1.9% | 100% | 100% | 1.1e-11 | 9.0e-8 | 0.16/2.0 |

**Table S2:** Complete solver performance on the weakly coherent snapshot workload (23,808 LPs). A looser tolerance ($\varepsilon=10^{-4}$) yields minimal improvements for first-order GPU solvers, allowing cuOpt PDLP to reach 0.34 LP/s and MPAX to solve 22% of LPs.

| Problem | m | n | Distinct bases | Infeasible | Simplex solves | Certified | manylp CPU LP/s | manylp GPU LP/s | HiGHS x32 LP/s | Speed-up | Status agreement | Max obj. error | Errors HiGHS / manylp |
|---|---|---|---|---|---|---|---|---|---|---|---|---|---|
| adlittle | 56 | 97 | 512+ | 85 | 957 | 100% | 414 | 402 | 795 | 0.52x | 100% | 7.3e-11 | 0/0 |
| afiro | 27 | 32 | 3 | 0 | 7 | 100% | 53,608 | 35,048 | 712 | 75.3x | 100% | 5.1e-16 | 0/0 |
| agg | 488 | 163 | 10 | 1,014 | 23 | 100% | 1,673 | 5,461 | 4,082 | 1.34x | 100% | 1.2e-14 | 0/0 |
| agg2 | 516 | 302 | 512+ | 0 | 917 | 100% | 389 | 608 | 7,125 | 0.09x | 100% | 2.6e-15 | 0/0 |
| agg3 | 516 | 302 | 512+ | 0 | 1,024 | 100% | 2,855 | 2,959 | 8,410 | 0.35x | 100% | 1.3e-14 | 0/0 |
| bandm | 305 | 472 | 512+ | 0 | 1,024 | 100% | 2,758 | 3,031 | 5,524 | 0.55x | 100% | 2.5e-12 | 0/0 |
| blend | 74 | 83 | 13 | 0 | 23 | 100% | 12,861 | 9,731 | 682 | 18.9x | 100% | 3.9e-14 | 0/0 |
| bnl1 | 643 | 1,175 | 476 | 541 | 599 | 99.2% | 564 | 401 | 586 | 0.96x | 100% | 6.7e-11 | 12/7 |
| boeing1 | 351 | 384 | 105 | 919 | 112 | 100% | 4,671 | 4,775 | 2,440 | 1.96x | 100% | 3.7e-15 | 0/0 |
| boeing2 | 166 | 143 | 236 | 520 | 237 | 100% | 1,397 | 1,173 | 16,280 | 0.09x | 100% | 2.2e-12 | 0/0 |
| brandy | 220 | 249 | 512+ | 0 | 1,024 | 100% | 3,954 | 3,719 | 7,947 | 0.50x | 100% | 5.3e-13 | 0/0 |
| capri | 271 | 353 | 512+ | 17 | 1,024 | 100% | 3,305 | 3,900 | 9,099 | 0.43x | 100% | 2.0e-14 | 0/0 |
| degen2 | 444 | 534 | 0 | 1,024 | 3 | 100% | 12,629 | 11,853 | 1,060 | 11.9x | 100% | 0 | 0/0 |
| e226 | 223 | 282 | 512+ | 5 | 1,024 | 100% | 4,407 | 4,154 | 8,617 | 0.51x | 100% | 7.3e-12 | 0/0 |
| etamacro | 400 | 688 | 415 | 0 | 417 | 100% | 440 | 466 | 7,069 | 0.07x | 100% | 3.2e-11 | 0/0 |
| finnis | 497 | 614 | 0 | 1,024 | 15 | 100% | 11,281 | 11,733 | 4,551 | 2.58x | 100% | 0 | 0/0 |
| fit1d | 24 | 1,026 | 1 | 0 | 1 | 100% | 87,618 | 81,752 | 9,066 | 9.66x | 100% | 0 | 0/0 |
| grow15 | 300 | 645 | 1 | 0 | 1 | 100% | 18,529 | 18,576 | 5,411 | 3.43x | 100% | 1.4e-16 | 0/0 |
| grow7 | 140 | 301 | 1 | 0 | 1 | 100% | 50,072 | 48,315 | 11,461 | 4.37x | 100% | 1.6e-16 | 0/0 |
| israel | 174 | 142 | 512+ | 0 | 1,023 | 100% | 4,146 | 4,604 | 10,156 | 0.45x | 100% | 3.0e-13 | 0/0 |
| kb2 | 43 | 41 | 1 | 0 | 1 | 100% | 195,329 | 149,731 | 672 | 291x | 100% | 2.0e-13 | 0/0 |
| lotfi | 153 | 308 | 512+ | 0 | 1,024 | 100% | 4,765 | 5,222 | 15,249 | 0.34x | 100% | 2.0e-14 | 0/0 |
| recipe | 91 | 180 | 1 | 0 | 1 | 100% | 138,514 | 120,642 | 887 | 156x | 100% | 2.1e-16 | 0/0 |
| sc105 | 105 | 103 | 4 | 0 | 7 | 100% | 28,024 | 24,106 | 867 | 32.3x | 100% | 2.0e-15 | 0/0 |
| sc205 | 205 | 203 | 4 | 0 | 6 | 100% | 26,069 | 21,904 | 747 | 34.9x | 100% | 1.3e-15 | 0/0 |
| sc50a | 50 | 48 | 3 | 0 | 7 | 100% | 38,528 | 30,894 | 858 | 44.9x | 100% | 1.7e-15 | 0/0 |
| sc50b | 50 | 48 | 1 | 0 | 1 | 100% | 189,711 | 144,962 | 810 | 234x | 100% | 1.5e-15 | 0/0 |
| scagr25 | 471 | 500 | 512+ | 0 | 1,023 | 100% | 2,546 | 2,888 | 7,803 | 0.37x | 100% | 1.2e-15 | 0/0 |
| scagr7 | 129 | 140 | 358 | 0 | 376 | 100% | 835 | 600 | 649 | 1.29x | 100% | 1.4e-15 | 0/0 |
| scfxm1 | 330 | 457 | 512+ | 0 | 857 | 100% | 824 | 931 | 8,709 | 0.11x | 100% | 2.0e-13 | 0/0 |
| scsd1 | 77 | 760 | 1 | 0 | 1 | 100% | 97,275 | 78,179 | 16,726 | 5.82x | 100% | 1.1e-14 | 0/0 |
| scsd6 | 147 | 1,350 | 7 | 0 | 10 | 100% | 13,257 | 11,498 | 7,647 | 1.73x | 100% | 2.0e-14 | 0/0 |
| scsd8 | 397 | 2,750 | 62 | 0 | 65 | 100% | 925 | 1,608 | 3,268 | 0.49x | 100% | 3.4e-14 | 0/0 |
| sctap1 | 300 | 480 | 512+ | 0 | 1,024 | 100% | 3,816 | 4,129 | 8,821 | 0.47x | 100% | 3.7e-15 | 0/0 |
| sctap2 | 1,090 | 1,880 | 512+ | 0 | 1,024 | 100% | 1,762 | 1,944 | 4,442 | 0.44x | 100% | 4.0e-16 | 0/0 |
| share1b | 117 | 225 | 308 | 0 | 349 | 100% | 849 | 657 | 14,150 | 0.06x | 100% | 1.7e-13 | 0/0 |
| share2b | 96 | 79 | 15 | 0 | 19 | 100% | 13,251 | 11,386 | 765 | 17.3x | 100% | 5.3e-13 | 0/0 |
| ship04l | 402 | 2,118 | 151 | 2 | 174 | 100% | 749 | 725 | 6,000 | 0.12x | 100% | 1.3e-15 | 0/0 |
| ship04s | 402 | 1,458 | 107 | 2 | 139 | 100% | 991 | 970 | 7,453 | 0.13x | 100% | 1.2e-15 | 0/0 |
| ship08s | 778 | 2,387 | 269 | 0 | 285 | 100% | 383 | 370 | 3,949 | 0.10x | 100% | 1.2e-15 | 0/0 |
| stair | 356 | 467 | 512+ | 0 | 1,024 | 100% | 1,713 | 1,886 | 2,680 | 0.70x | 100% | 5.0e-12 | 0/0 |
| standata | 359 | 1,075 | 1 | 722 | 3 | 100% | 33,062 | 30,859 | 13,503 | 2.45x | 100% | 0 | 0/0 |
| stocfor1 | 117 | 111 | 1 | 0 | 1 | 100% | 81,387 | 76,729 | 993 | 82.0x | 100% | 3.3e-15 | 0/0 |
| vtp.base | 198 | 203 | 3 | 1,021 | 18 | 100% | 19,647 | 19,474 | 7,269 | 2.70x | 100% | 4.5e-15 | 0/0 |

**Table S3:** Performance across 44 Netlib LPs (1,024 scenarios per problem). Distinct bases indicate the number of optimal bases required to certify the feasible scenarios (512+ denotes the pool capacity was reached).